\documentclass[apj,twocolumn]{openjournal}
\usepackage{graphicx}
\usepackage{caption}
\usepackage{subcaption}
\usepackage{xcolor}
\usepackage{amsmath}
\usepackage{amsfonts}
\usepackage{amssymb}
\usepackage{upgreek}
\usepackage[pdfborder={0 0 0}]{hyperref}
\usepackage{txfonts}
\usepackage{lipsum} 
\usepackage[utf8]{inputenc}
\usepackage{upgreek}
\usepackage{multirow}
\usepackage{booktabs}
\usepackage{hyperref}
\usepackage{cleveref}
\newcommand{\orcidauthor}[3]{\author{\href{http://orcid.org/#1}{#2 \openin1 Orcid-ID.png \ifeof1 \else \hskip2pt\includegraphics[width=9pt]{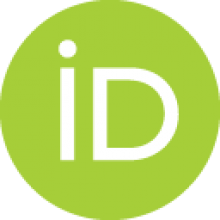}\fi}$^{#3}$}}

\newcommand{\Tsys}{T_{\text{sys}}}
\newcommand{\Tsky}{T_{\text{sky}}}

\newcommand{\dm}{$\text{pc~cm}^{-3}$}

\DeclareUnicodeCharacter{2212}{-}

\begin{document}

\title{Radio Observations of a Candidate Redback Millisecond Pulsar: 1FGL J0523.5$-$2529}

\orcidauthor{0000-0002-5927-0481}{O.~A.~Johnson}{1,2}
\orcidauthor{0000-0002-4553-655X}{E.~F.~Keane}{1}
\orcidauthor{0000-0001-7185-1310}{D.~J.~McKenna}{3}
\orcidauthor{0000-0002-9586-7904}{H.~Qiu}{4}
\orcidauthor{0000-0003-1699-8867}{S.~J.~Swihart}{5}
\orcidauthor{0000-0002-1468-9668}{J.~Strader}{6}
\orcidauthor{0000-0001-7697-7422}{M.~McLaughlin}{7}

\affiliation{$^1$ School of Physics, Trinity College Dublin, College Green, Dublin 2, D02 PN40, Ireland}
\affiliation{$^2$ Radio Astronomy Laboratory, University of California, Berkeley, CA, USA}
\affiliation{$^3$ ASTRON, The Netherlands Institute for Radio Astronomy, Oude Hoogeveensedijk 4, 7991 PD Dwingeloo, The Netherlands}
\affiliation{$^4$ SKA Observatory, Jodrell Bank, Lower Withington, Macclesfield SK11 9FT, UK}
\affiliation{$^5$ Science, Systems and Sustainment Division, Institute for Defense Analyses, Alexandria, VA 22305, USA}
\affiliation{$^6$ Center for Data Intensive and Time Domain Astronomy, Department of Physics and Astronomy, Michigan State University, East Lansing, MI 48824, USA}
\affiliation{$^7$ Department of Physics and Astronomy, West Virginia University, Morgantown WV 26501, USA}


\begin{abstract}
{Redback pulsars are a subclass of millisecond pulsar system with a low-mass non-degenerate companion star being ablated by the pulsar. They are of interest due to the insights they can provide for late-stage pulsar evolution during the recycling process. J0523.5-2529 is one such candidate where redback-like emission has been seen at multiple wavelengths except radio. It is a system with a binary orbit of 16.5 hours and a low-mass non-degenerate companion of $\sim0.8~\text{M}_\odot$.}
{The aim of this work was to conduct follow-up radio observations to search for any exhibited radio pulsar emission from J0523.5$-$2529.} 
{This work employs a periodicity and single burst search across 74\% of the system's orbital phase using a total of 34.5 hours of observations. Observations were carried out using the Murriyang Telescope at Parkes and the Robert C. Byrd Green Bank Telescope (GBT).}
{Despite extensive orbital phase coverage, no periodic or single-pulse radio emission was detected above a signal-to-noise threshold of 7. }
{A comprehensive search for radio pulsations from J0523.5-2529 using Parkes and GBT yielded no significant emission, likely due to intrinsic faintness, scattering, or eclipses by the companion’s outflow. The results demonstrate the elusiveness of the pulsar component in some redback systems and highlight the need for multi-wavelength follow-up and higher-frequency radio observations to constrain the source nature and binary dynamics.}
\end{abstract}
\maketitle


\section{Introduction}
Millisecond pulsars (MSPs) are rapidly rotating neutron stars with spin periods typically below 10 milliseconds. They are believed to be `recycled'~\citep{bh91}, spun up through accretion of matter from a binary companion, which transfers angular momentum and accelerates their rotation. Due to their stable rotation, MSPs are valuable tools for many astrophysical studies, including tests of strong-field gravity~\citep{kramer2021} and the study of low-frequency gravitational waves~\cite{fb90}.

Redbacks are a subclass of binary millisecond pulsars, with a non-degenerate low-mass companion ($\gtrsim 0.1~ \text{M}_\odot$) that exhibit significant mass transfer or ablation of the companion star due to the pulsar wind \citep{roberts_x-ray_2017}. This can result in observable flares, eclipses, and temperature variations on the star’s surface, leading to emission across the entire electromagnetic spectrum \citep{papitto_transitional_2022,kirichenko_black_2024}. Some redbacks belong to the ‘transitional MSP’ subclass, which oscillates between an accretion-powered X-ray pulsar phase and a rotation-powered radio MSP phase, due to changes in the mass transfer rate \citep{Papitto2013, papitto_transitional_2022}, to date three such redbacks have been observed in changing transition states, PSRs~J1023$+$0038 \citep{Stappers2014}, J1824-2452I \citep{Papitto2013}, and J1227-4853 \citep{Roy2015}. 

Redbacks are thought to be a distinct evolutionary outcome of binary millisecond pulsar systems, likely shaped by the efficiency of evaporation and irradiation feedback~\citep{chen2013, devito2020}. In these systems, the neutron star has already been spun up by accretion from its companion~\citep{th06}, but the mass transfer is often unstable, leading to strong pulsar winds that partially or completely ablate the companion star. The ablated material can also cause radio eclipses through scattering and absorption~\citep{broderick_low-radio-frequency_2016}. The long-term fate of redbacks remains uncertain: they may persist for Gyr timescales, evolve into isolated MSPs by fully ablating their companion, or eventually become fully recycled MSPs with a low-mass white dwarf companion~\citep{chen2013, zanon_evidence_2021}.

1FGL J0523.5$-$2529 (herein J0523) was first discovered as a $\gamma$-ray source with the \textit{Fermi} Large Area Telescope \citep[LAT;][]{abdo_second_2013} and flagged as a candidate MSP by \citet{ackermann_statistical_2012}. Follow-up optical observations by \citet{strader_1fgl_2014} determined the presence of a low-mass companion star with a lower limit of $0.8~\text{M}_\odot$ and periodic modulation of $16.5$~hr. The inferred compact primary from the optical radial velocities and mass ratio, combined with the presence of a low-mass companion, suggests that the source is potentially a redback. Additional supporting evidence for this includes its gamma-ray and X-ray luminosity and location outside the Galactic plane. Further X-ray and optical observations by \citet{halpern_luminous_2022} revealed episodic flaring in both X-rays and at optical wavelengths, with peak luminosities up to $100$ times the quiescent state. Moreover, J0523 exhibited broad luminous Balmer and He~I emission which has also been observed in other spider systems \citep{strader_1fgl_2014}.

J0523 has a more massive companion compared to other known redbacks, which generally lie in the range of $\sim 0.1 - 0.5~\text{M}\odot$ \citep{ATNFCat_2005,smedley_formation_2015}.  The high companion mass ($\sim 0.8 \text{M}\odot$) is unique to the currently known redback population which ranges from $0.10-0.44~\text{M}\odot$ \citep{keane_survey_2018, koljonen_spidercat_2025}, this suggests that J0523 may represent an earlier stage in redback evolution. Alternatively it could be a system that forms with a higher initial mass companion with shorter orbital periods, evolving more rapidly. It has also shown ellipsoidal modulation that exhibits asymmetry in its optical lightcurve caused by the companion's tidal deformation \citep{strader_1fgl_2014}, along with temperature variations on the surface of the companion star \citep{halpern_luminous_2022, strader_1fgl_2014}. These variations may be in part due to pulsar-driven heating. J0523 presents an interesting case to conduct multi-wavelength investigations of pulsar-companion interactions and the evolutionary pathways of redback systems. To date, there has been no radio emission detected from J0523 despite radio observations from MeerKAT \citep{thongmeearkom_targeted_2024} and observations from GBT at 820~MHz \citep{SanpaArsa2016PhD}. The lack of detected radio emission could be due to absorption or scattering by ionized material from the companion, a common feature in redback systems \citep{zic_discovery_2024}. 

The Gaia Data Release 3 (GDR3) contain parallax measurements for J0523 which give a distance of $2.2 \pm 0.21~\text{kpc}$ implying a dispersion measure (DM) of $\sim 40$ or $\sim 33$~\dm according to either the \textsc{NE2001}~\citep{cordes_ne2001i_2003} or \textsc{YMW16}~\citep{yao_new_2017} electron density models. The errors are taken to be the difference in DM value given by the upper and lower distance measurements from the stated value.  

The structure of this paper is as follows §\ref{sec:obs} describes the observations, §\ref{sec:methods} outlines the data reduction and analysis, §\ref{sec:results} presents the results, and §5 concludes the findings.

\section{Observations} \label{sec:obs}

\begin{figure}
    \centering
    \includegraphics[width=0.99\linewidth]{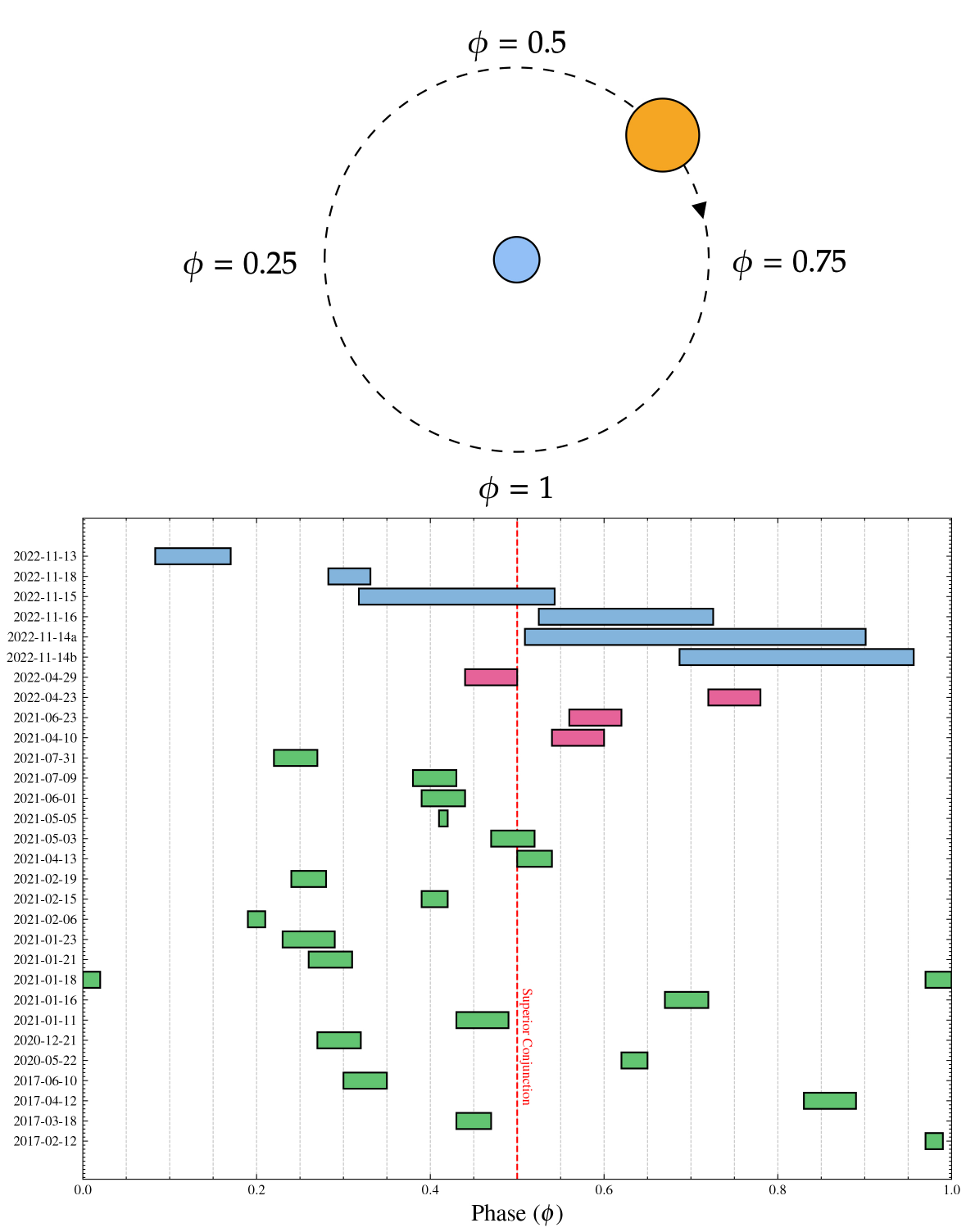}
    \caption{\textit{Top:} Phase convention used during this search, with superior conjunction at $\phi = 0.5$. Pulsar shown in blue, companion shown in orange (top down view). \textit{Bottom:}  Bar chart of epochs and corresponding phase coverage of the observation campaign from Parkes (Blue), MeerKAT (Pink) and GBT (Green). }
    \label{Fig.:gantt_chart}
\end{figure}

Observations of J0523 were carried out using the ultra-wide-bandwidth, low-frequency receiver (UWL; 704 -- 4032~MHz) on the Murriyang telescope at Parkes~\citep{hobbs_ultra-wide_2020} and with the Prime Focus 1 (PF1; 680 -- 920~MHz), L-band (1150 -- 1730~MHz) and S-Band (1730--2600~MHz) receivers on the Robert C. Byrd Green Bank Telescope (GBT) at the Green Bank Observatory. The observing epochs were selected to ensure coverage across all orbital phases, with additional emphasis on superior conjunction (using a phase convention where this occurs at $\phi = 0.5$) where potential eclipses due to ionized material from the companion are least likely to occur, this as illustrated in Fig. \ref{Fig.:gantt_chart}.  The Parkes observations were recorded using the Medusa backend \citep{hobbs_ultra-wide_2020}, with a total bandwidth of $3.3$~GHz, a frequency resolution of $0.25$~MHz and a time resolution of $64~\upmu\text{s}$ under project code PX094\footnote{\scriptsize{\texttt{https://www.parkes.atnf.csiro.au/observing/schedules/PX\_numbers.php}}}. The GBT observations were recorded using the Vegas backend, with varying bandwidths (see. Table. \ref{tab:combined_observations}), with frequency resolutions as follows for the GBT observations, $0.12$~MHz for the PF1 observations, $0.14$~MHz for L-band and $0.21$~MHz for S-band, and with time resolution of $82~\upmu\text{s}$ across the three bands. These observations were taken under project codes AGBT17A-102, AGBT20A0-538 and AGBT20A-202. Some of the GBT observations were pointed at incorrect co-ordinates\footnote{See supplementary Table. for more details on all GBT pointings.}; these pointings were also searched using the same outlined methods and served as an additional test of our pipelines (see below). Details of each observation can be found in Table \ref{tab:combined_observations}. 

The ephemerides in Table~\ref{tab:ephemeris_values} were derived from optical radial velocity measurements by \cite{strader_1fgl_2014}. Although the ephemerides in Table~\ref{tab:ephemeris_values} are based on data from over a decade ago, the orbit remains phase coherent over the full timespan of the observations. This indicates that the pulsar's spin and orbital parameters have remained stable enough for accurate phase predictions \citep{strader_1fgl_2014,halpern_luminous_2022}. The orbital solution from the optical radial velocities in \citet{strader_1fgl_2014} yields a small but non-zero eccentricity (\(e = 0.0040(6)\); see Table~\ref{tab:ephemeris_values}), which is likely not physical and instead reflects heating-induced asymmetries that shift the measured centre of light away from the centre of mass \citep{halpern_luminous_2022}. A circular orbit is assumed in the analysis.

\begin{table*}
	\centering
	\caption{Ephemeris values of note for J0523$-$2529. All values are taken from \cite{strader_1fgl_2014} and \cite{halpern_luminous_2022}.}
	\label{tab:ephemeris_values}
	\begin{tabular}{ccccccccc} 
		\hline
		RA & DEC & $P_{\text{orb}}$ & $k_2$ & $a \sin i$ & $e$ & $q$ & $d$ & $\text{MJD}_{\phi = 0.5}$ \\
        (J2000) & (J2000) & (days) & ($\text{km\,s}^{-1}$) & (s) & & & (kpc) &  \\ \hline 
        05:23:17.18 & $-$25:27:37.4 & $0.68813(28)$ & 190.3(11) & 0.359 & $0.0040(6)$ & $0.6(6)$ & $2.2(21)$ & $56577.14636(37)$ \\
		\hline
	\end{tabular}
\end{table*}

\begin{table*}
    \caption{Details of the search mode observations taken with the Parkes UWL (Project Code: PX094), GBT (Project Codes: AGBT17A-102, AGBT20A-538 and AGBT20A-202) and  TRAPUM \citep{thongmeearkom_targeted_2024}.}
    \centering
    \footnotesize
    \begin{tabular}{lccccc}
        \toprule
        \multicolumn{6}{c}{\textbf{Parkes (UWL)}} \\
        \cmidrule(r){1-6}
        MJD & Obs. Date (UT) & $t_{\text{obs}}$ (hrs) & Phase Range $\phi$ & Frequency (GHz) & \\
        \midrule
        59892.4866 & 2022-11-09 11:40 & 6.5  & 0.40 -- 0.79 & 0.7 -- 4.0 \\
        59896.5251 & 2022-11-13 12:36 & 1.4  & 0.27 -- 0.35 & 0.7 -- 4.0 \\
        59897.4790 & 2022-11-14 11:29 & 4.4  & 0.65 -- 0.92 & 0.7 -- 4.0 \\
        59898.6677 & 2022-11-15 16:01 & 3.7  & 0.38 -- 0.61 & 0.7 -- 4.0 \\
        59899.4708 & 2022-11-16 11:18 & 3.3  & 0.55 -- 0.75 & 0.7 -- 4.0 \\
        59901.4666 & 2022-11-18 11:11 & 0.8  & 0.45 -- 0.50 & 0.7 -- 4.0 \\
        \midrule
        \multicolumn{6}{c}{\textbf{GBT}} \\
        \cmidrule(r){1-6}
        MJD & Obs. Date (UT) & $t_{\text{obs}}$ (hrs) & Phase Range $\phi$ & Frequency (GHz) & \\
        \midrule
        57796.1495 & 2017-02-12 03:35 & 0.4 & 0.97 -- 0.99 & 0.68 -- 0.92 \\
        57830.8724 & 2017-03-18 20:56 & 0.8 & 0.43 -- 0.47 & 0.68 -- 0.92 \\
        57855.9199 & 2017-04-12 22:04 & 1.0 & 0.83 -- 0.89 & 0.68 -- 0.92 \\
        57914.7364 & 2017-06-10 17:40 & 0.8 & 0.30 -- 0.35 & 0.68 -- 0.92 \\
        58991.8799 & 2020-05-22 21:07 & 0.4 & 0.62 -- 0.65 & 1.73 -- 2.60 \\
        59204.2670 & 2020-12-21 06:24 & 0.8 & 0.27 -- 0.32 & 1.73 -- 2.60 \\
        59225.0261 & 2021-01-11 00:37 & 0.9 & 0.43 -- 0.49 & 1.73 -- 2.60 \\
        59230.0052 & 2021-01-16 00:07 & 0.9 & 0.67 -- 0.72 & 1.73 -- 2.60 \\
        59232.9653 & 2021-01-18 23:10 & 0.8 & 0.97 -- 0.02 & 1.73 -- 2.60 \\
        59235.2297 & 2021-01-21 05:30 & 0.8 & 0.26 -- 0.31 & 1.73 -- 2.60 \\
        59237.9603 & 2021-01-23 23:02 & 1.0 & 0.23 -- 0.29 & 1.73 -- 2.60 \\
        59251.0070 & 2021-02-06 00:10 & 0.3 & 0.19 -- 0.21 & 1.73 -- 2.60 \\
        59260.0882 & 2021-02-15 02:06 & 0.6 & 0.39 -- 0.42 & 1.73 -- 2.60 \\
        59264.1177 & 2021-02-19 02:49 & 0.7 & 0.24 -- 0.28 & 1.73 -- 2.60 \\
        59317.9707 & 2021-04-13 23:17 & 0.7 & 0.50 -- 0.54 & 1.73 -- 2.60 \\
        59337.9066 & 2021-05-03 21:45 & 0.7 & 0.47 -- 0.52 & 1.15 -- 1.73 \\
        59339.9243 & 2021-05-05 22:11 & 0.3 & 0.41 -- 0.42 & 1.73 -- 2.60 \\
        59366.7467 & 2021-06-01 17:55 & 0.9 & 0.39 -- 0.44 & 1.73 -- 2.60 \\
        59404.5904 & 2021-07-09 14:10 & 0.8 & 0.38 -- 0.43 & 1.73 -- 2.60 \\
        59426.5008 & 2021-07-31 12:01 & 0.7 & 0.22 -- 0.27 & 1.73 -- 2.60 \\
        \midrule
        \multicolumn{6}{c}{\textbf{MeerKAT (TRAPUM)}} \\
        \cmidrule(r){1-6}
        MJD & Obs. Date (UT) & $t_{\text{obs}}$ (hrs) & Phase Range $\phi$ & Frequency (GHz) & \\
        \midrule
        59314.5519 & 2021-04-10 13:14 & 1.0 & 0.54 -- 0.60 & 0.9 -- 1.7 \\
        59388.2008 & 2021-06-23 04:49 & 1.0 & 0.56 -- 0.62 & 0.6 -- 1.0 \\
        59692.4609 & 2022-04-23 11:03 & 1.0 & 0.72 -- 0.78 & 0.6 -- 1.0 \\
        59698.4614 & 2022-04-29 11:04 & 1.0 & 0.44 -- 0.50 & 0.9 -- 1.7 \\
        \bottomrule
    \end{tabular}
    \label{tab:combined_observations}
\end{table*}

\section{Methods} \label{sec:methods}

\begin{figure}
    \centering
    \includegraphics[width=0.99\linewidth]{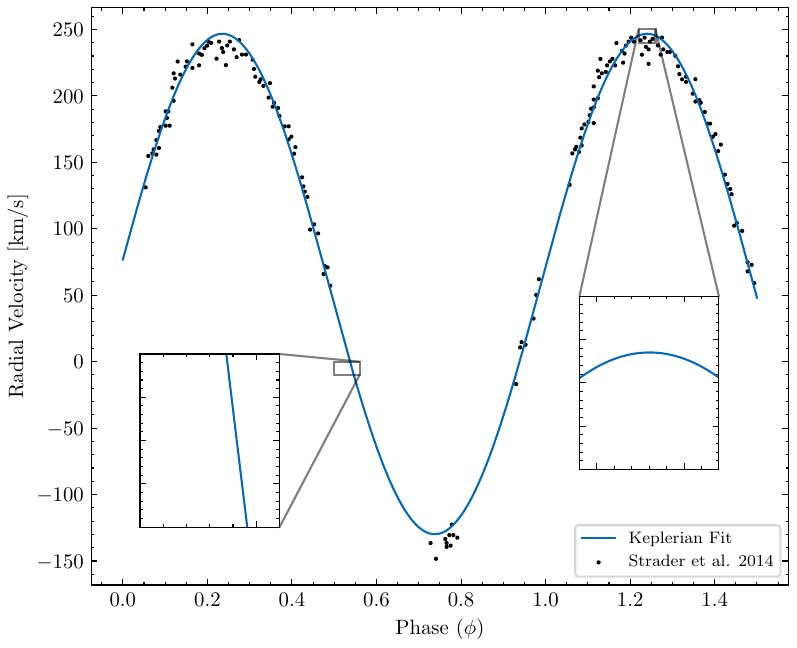}
    \caption{Radial velocity measurements from \cite{strader_1fgl_2014} as a function of phase. This illustrates the need to separate observations into chunks of $\lesssim P_b/10$ in duration so as to maintain an approximately linear radial acceleration for periodicity searches. }
    \label{Fig.:radial-velocity}
\end{figure}

The collected UWL data were split into three segments (704 -- 1856 MHz; 1856 -- 3008 MHz; 3008 -- 4032 MHz). Of these channels 15.3\% were masked due to persistent and strong radio frequency interference (RFI; \cite{zhou_ultra_wide_2022}). Following this, \textsc{PRESTO} \citep{ransom_presto_2011} was used to further mitigate RFI and de-disperse the data in a range from zero to 60~\dm so as to cover and exceed the expected DM of any putative radio pulsar given the distance ($2.2\pm0.2~\text{pc}$) and location in the Galactic plane ($l=228.17^\circ$, $b=-29.83^\circ$). The same method was applied to the GBT data but the spliting in frequency was omitted as the individual fractional bandwidths were significantly smaller than those of the UWL.  As illustrated in \cref{Fig.:radial-velocity} to maintain an approximate linear acceleration, each of the observations that exceeded $P_b \leq 10$ was split such that observations were all between 60 and 100 minutes, depending on the total length of the observation track. 

\begin{figure}
    \centering
    \includegraphics[width=0.99\linewidth]{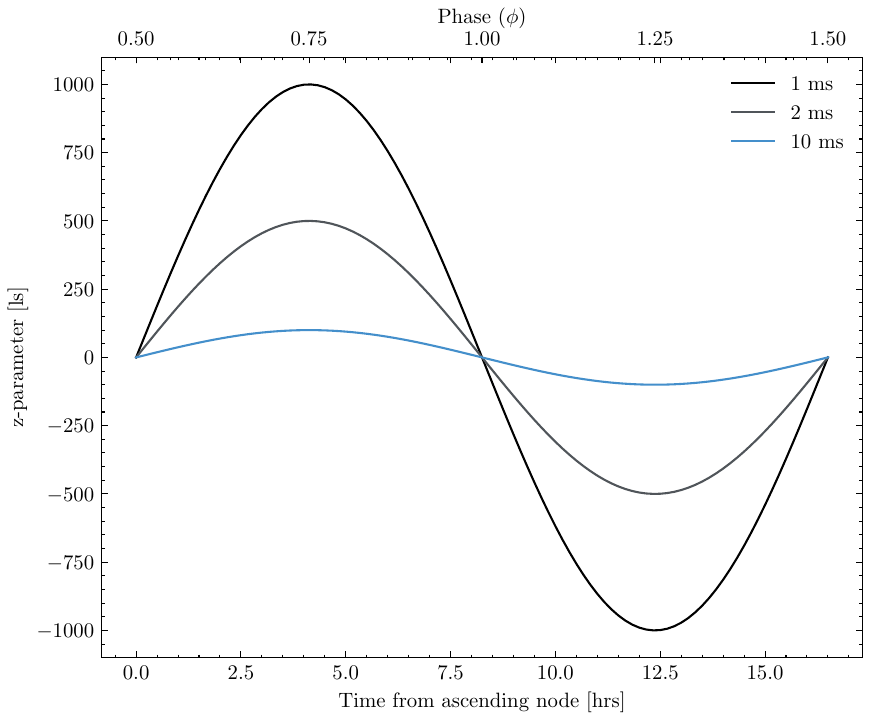}
    \caption{Z-parameter values for various J0523 trial spin periods. Here the max z-value corresponds to the value at the time from ascending node.}
    \label{Fig.:z_plot}
\end{figure}

As J0523 is in a binary, the system’s motion will cause the pulse frequency to smear across Fourier bins \citep{Ransom-z-2001}. To mitigate this effect, the dedispersed time-series must be divided into smaller time intervals where the radial velocity of the system can be assumed constant as shown in Fig. \ref{Fig.:radial-velocity}. The Z-parameter, which represents the projected semi-major axis of the pulsar's orbit in light-seconds, is given by:
\begin{equation}
    z \simeq \dot f T^2_\text{obs}\;,
    \label{eq:z-parameter}
\end{equation}
where $T_{\text{obs}}$ is the observation duration and $\dot f$ is the pulsars apparent spin-down which is given by: 
\begin{equation}
    \dot f = f_{\text{spin}} \frac{4 \pi^2 a \sin i}{cP^2_B} \sin \left[ \frac{2\pi (t - t_{\text{asc}})}{P_B} \right]\;,
    \label{eq:spindown}
\end{equation}
where $P_B$ is the system’s orbital period. In this expression, $f_{\text{spin}}$ is the intrinsic spin frequency of the pulsar, $a$ is the semi-major axis of the pulsar's orbit, $i$ is the orbital inclination angle, $c$ is the speed of light, $t$ is the time of observation, and $t_{\text{asc}}$ is the time of ascending node (i.e. when the pulsar crosses the plane of the sky moving away from the observer, $\phi=0.25$). Using the values listed in Table~\ref{tab:ephemeris_values}, along with Eq.\ref{eq:z-parameter} and Eq.\ref{eq:spindown}, the estimated values for the Z-parameter were calculated and are presented in Fig.~\ref{Fig.:z_plot}. An acceleration search was carried out using \textsc{PRESTO} \citep{ransom_presto_2011} along with a single pulse search using both \textsc{PRESTO} and \textsc{TransientX} \citep{men_transientx_2024}. 

An already known source, PSR~J0520$-$2553~\citep{lml+98} fell within the beam of one of the S-band GBT observations and was detected by both the single pulse and acceleration search pipeline, providing validation for the developed methodology (see appendix \ref{app:J0520}).

\section{Results} \label{sec:results}
No significant pulsations were detected from the observation campaign. The minimum detectable flux density per channel, $( S_{\text{min,chan}}, \text{in Jy})$ was calculated using Eq. \ref{eq:radiometer_equation} which is an adapted form of the radiometer equation~\citep{pulsar_handbook}: 

\begin{equation}
\begin{split}
S_{\min,\rm{chan}}(f, l, b) &= \frac{\left[ T_{\text{sky}}(l,b) + T_{\text{int}}(f) \right] \cdot 2 k_B}{A_{\text{eff}}(f)} \cdot \beta \\ & \times \frac{\text{S/N}_{\min}}{\sqrt{n_p\, t_{\text{obs}}\, B_{\text{channel}}}} \cdot \sqrt{\frac{W}{P - W}}
 \label{eq:radiometer_equation}
\end{split}
\end{equation}

Here, $f$ is the observing frequency, $T_{\text{sky}}(l, b)$ is the sky temperature at Galactic coordinates $(l, b)$, and $T_{\text{int}}(f)$ is the internal system temperature. $A_{\text{eff}}(f)$ is the effective collecting area of the telescope, $k_B$ is the Boltzmann constant, and $\beta$ is a factor accounting for system losses such as digitisation. $n_p$ is the number of summed polarisations and $B_{\text{channel}}$ is the width of an individual frequency channel. $P$ is the pulsar period and $W$ is the effective observed pulse width. Each individual channel $S_{\text{min}}$ is calculated with a boolean mask applied to account for RFI contamination, which is taken to be averages found in the band. The collective $S_\text{min,chan}$ for each observation is calculated using quadrature method as shown below:  

\begin{equation}
    S_{\min}(f, l, b) = \left[ \sum_f \frac{1}{S_{\min}(f, l, b)^2} \right]^{-1/2}
    \label{eq:quadrature}
\end{equation}

\begin{figure*}
    \centering
    \begin{subfigure}{0.49\linewidth}
        \centering
        \includegraphics[width=\linewidth]{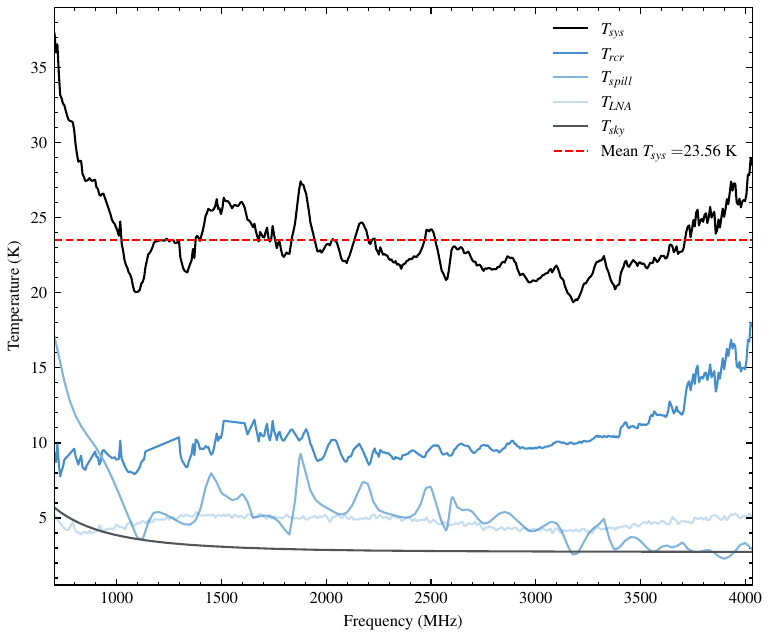}
        \label{Fig.:tsys_parkes}
    \end{subfigure}
    \hfill
    \begin{subfigure}{0.49\linewidth}
        \centering
        \includegraphics[width=\linewidth]{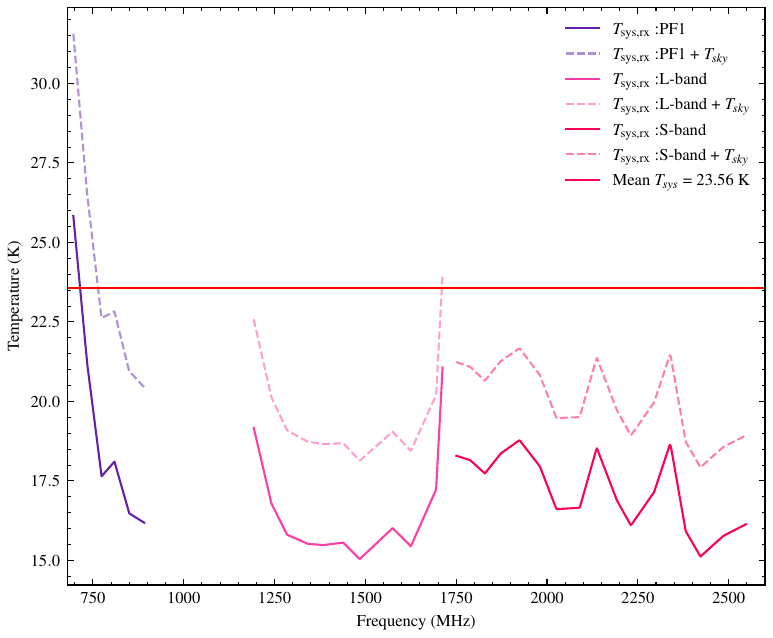}
        \label{Fig.:tsys_gbt}
    \end{subfigure}
    \caption{System temperatures across both Parkes \textit{(left)} and GBT \textit{(right)} bands. The plots show the contributions from the sky, Low-Noise Amplifier (LNA), spillover, and receiver temperatures, along with the total system temperature. Sky temperature values were obtained from \citet{zheng_improved_2017}, Parkes values were referenced from \citet{hobbs_ultra-wide_2020}, and GBT values from the proposer's guide.}
    \label{Fig.:tsys}
\end{figure*}

All values in the calculation for the UWL $\Tsys$ were directly extracted from \cite{hobbs_ultra-wide_2020} (see. Fig \ref{Fig.:tsys}) with the exception of $\Tsky$ which is target-dependent where the \cite{zheng_improved_2017} sky model was used. For GBT values were taken from the proposers guide\footnote{\url{https://www.gb.nrao.edu/scienceDocs/GBTpg.pdf}} and also shown in Fig. \ref{Fig.:tsys}.

\begin{table*}
\centering
\caption{Upper limit flux densities (\( S_{\min} \)) for a 10\% duty cycle set by Parkes and GBT observations. A \( t_\text{obs} \) of 1 hour and the fractional bandwidth of each frequency slice is used. The average RFI mask for the respective band throughout the observation campaign is also taken into account in the calculations.}
\label{tab:smin_freqs}
\begin{tabular}{lccccc}
\hline
Telescope & Center Frequency (MHz) & $T_{\mathrm{sys}}$ (K) & $A_{\mathrm{eff}}$ (m$^2$) & Bandwidth (MHz) & $S_{\min,\,10\%}$ (mJy) \\
\hline
\multirow{3}{*}{Parkes}
 & 1280 & 34.36 & 2052 & 1152 & 0.087 \\
 & 2432 & 25.10 & 1856 & 1152 & 0.034 \\
 & 3520 & 21.30 & 1919 & 1024 & 0.300 \\
\hline
\multirow{3}{*}{GBT}
 & 820  & 29.18 & 5548 & 240 & 0.022 \\
 & 1420 & 20.06 & 5606 & 580 & 0.012 \\
 & 2380 & 19.79 & 5649 & 860 & 0.010 \\
\hline
\end{tabular}
\end{table*}

The derived flux density limits (see Table. \ref{tab:smin_freqs}) constrain the possible pulsar emission at these frequencies, indicating that either J0523 is intrinsically faint, highly scattered, or possibly experiencing eclipsing effects from the companion's wind. At the distance of $2.2$~kpc, the most stringent upper limit on the pseudo-luminosity is $L_{1400} \lesssim 106~\mu\mathrm{Jy\,kpc}^2$, assuming a pulsar spectral index of $-1.6$ to scale from the GBT S-band limit of $S_{2000} < 0.010$~mJy (see Table. \ref{tab:smin_freqs}) \citep[][]{Jankowski2018}. This limit is below the measured $L_{1400}$ of most known redbacks, such as J1048+2339 \citep[$L_{1400} \sim 340~\mu\mathrm{Jy\,kpc}^2$,][]{Deneva2016} and J1431$-$4715 \citep[$L_{1400} \sim 1.8~\mathrm{mJy\,kpc}^2$,][]{Bates2015}, and comparable to the very faintest detected redbacks like J1957+2516 \citep[$L_{1400} \sim 140~\mu\mathrm{Jy\,kpc}^2$,][]{Stovall2016} and J1306$-$40 \citep[$L_{1400} \sim 100$--$200~\mu\mathrm{Jy\,kpc}^2$,][]{keane_survey_2018}, which has a relatively massive companion with an upper mass limit of $\sim 0.44~M_\odot$. The radio non-detection therefore probes pseudo-luminosities as low as or lower than those of the faintest confirmed redbacks, suggesting that J0523, if a redback, is either unusually radio-faint or is obscured by scattering or eclipses.

Interstellar scattering and dispersion effects were considered using the \citet{yao_new_2017} electron density model. For J0523’s estimated DM, pulse broadening at $1.4$~GHz is expected to be in the range of $0.1$--$0.5$~ms \citep{ATNFCat_2005}, which is a significant fraction of the expected pulse width for a millisecond pulsar and thus reduces sensitivity, particularly at the lower end of the UWL band and in the PF1 observations due to smearing.

No significant single bursts were found above a threshold S/N of 7, suggesting that J0523 does not emit detectable bright single pulses within these observation limits. As mentioned in §\ref{sec:obs} not all pointings under the GBT project codes targeted the correct sky position, in turn the pipeline serendipitously detected PSR~J0520$-$2553 using both the acceleration search and single-burst detection pipelines, as shown in Fig. \ref{Fig.:J0520-2553}. This detection provides verification of the pipelines functionality. An estimate of the flux density of J0520–2553 using S-band and the measured signal-to-noise ratio of the folded detection gives a flux density of $S_{2000} = 0.05~\text{mJy}$

Persistent RFI meant that 15.3\% of the total bandwidth for UWL observations were unusable at all times. Additional masking on top of this was also needed as at times there was a large amount of contamination from 704--1856~MHz during the observation campaign. In some cases in excess of 90\% of channels for large portions of the observing run were unusable. Given the significant RFI affecting the observations, future searches may benefit from utilizing alternative instruments and observatory sites with lower RFI contamination, e.g. the UHF band on MeerKAT. 

\section{Conclusions}
This study conducted a comprehensive search for radio pulsations from 1FGL~J0523.5$-$2529 using the Parkes and GBT telescopes. Performing periodicity and single burst searches, no significant radio emission was detected. The derived flux density upper limits and implied pseudo-luminosity limits constrain the radio emission properties of J0523, suggesting that it may be either an intrinsically faint or highly scattered pulsar, or that its emission is periodically eclipsed by material from the companion star. The lack of a detection aligns with previous non-detections of radio pulsations from this source, reinforcing the possibility that J0523 exhibits characteristics typical of eclipsing redbacks, where the pulsar wind interacts with the companion's outflow. The unusual properties of the companion star, including its unusually high mass and evidence for significant surface asymmetries, may contribute to extended or more opaque eclipses compared to typical redbacks. Additionally, significant levels of radio frequency interference (RFI) impacted portions of the available bandwidth in the Parkes observations, reducing overall sensitivity.

Further investigations using deep X-ray and archival gamma-ray observations could provide insights into J0523’s high-energy properties and possible timing, while continued radio monitoring at L-band or higher frequencies, may help confirm or rule out its pulsar nature. Multi-wavelength timing studies could also offer a more complete picture of its binary dynamics and evolutionary state. Ultimately improving observational constraints will be crucial in determining J0523's place within the spider population.

\section*{acknowledgements}
The authors acknowledge the insights and comments from Jules Halpern, Karen Perez, Wael Farah, Colin Clark, Scott Ransom, Lawrence Toomey, Joe Glaser and Yunpeng Men. OAJ acknowledges the support of the Trinty College Dublin, School of Physics and Breakthrough Listen which is managed by the Breakthrough Prize Foundation. This work was performed on the OzSTAR national facility at Swinburne University of Technology. The OzSTAR program receives funding in part from the Astronomy National Collaborative Research Infrastructure Strategy (NCRIS) allocation provided by the Australian Government.  JS acknowledges support from National Science Foundation grant AST-2205550. We also thank the anonymous reviewer for their careful reading of our manuscript and their many insightful comments and suggestions.


\bibliography{refs.bib}
\bibliographystyle{aasjournal}

\newpage
\section{appendix}
\subsection{Detection of J0520-2553} \label{app:J0520}

\begin{figure}[h!]
    \caption{Flux limits in mJy for S-band observations of J0520-2553 from}
    \centering
    \includegraphics[width=0.9\linewidth]{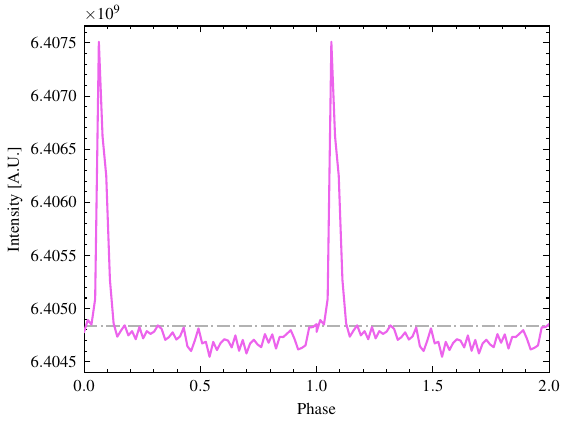}
    \includegraphics[width=0.9\linewidth]{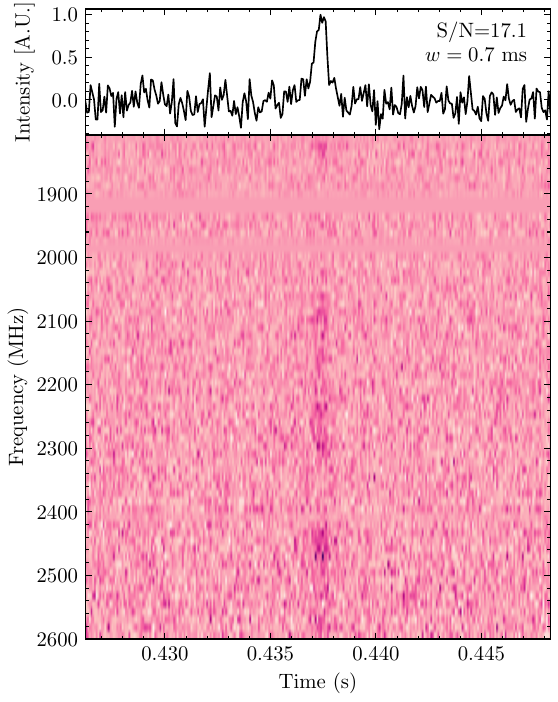}
    \caption{\textit{Top:} Pulse profile of J0520-2553 obtained from the accelsearch pipeline at a DM of 33.7~\dm. \textit{Bottom:} Single burst pulse also detected at 33.7~\dm~ in S-band.}
    \label{Fig.:J0520-2553}
\end{figure}

As stated in §\ref{sec:results} there was a serendipitous detection of J0520-2553 (see Fig. \ref{Fig.:J0520-2553}), using both the acceleration search and single burst search pipelines in S band. The fluxes were calculated using eq. \ref{eq:radiometer_equation} along with applying an offset from the boresight which is assumed to be Gaussian,  $\exp\left( -\left({4 \ln 2 \cdot \theta^2}\right)/{\text{FWHM}^2} \right)$. Where $\theta$ is the angular separation from the boresight. 

\begin{table}[h]
    \centering
    \caption{Flux limits in mJy for S-band observations of J0520-2553 attained from each of the search pipelines in this work.}
    \begin{tabular}{lcc}
        \hline
        Search Method & S/N & \(S_{2000}\) (mJy) \\
        \hline
        PRESTO Folding & 50.2  & 0.05 \\
        PRESTO Single Pulse & 16.6 & 164 \\
        TransientX Single Pulse & 17.2 & 170 \\
        \hline
    \end{tabular}
    \label{tab:J0520_fluxes}
\end{table}

\vspace{10pt} 

\end{document}